\documentclass[]{aa}

\usepackage{graphicx}
\usepackage{epsfig}
\usepackage{graphics}
\begin{document}%
   \title{Massive protostars as gamma-ray sources}

   \author{V. Bosch-Ramon\inst{1},
   G.~E. Romero\inst{2,3,}\thanks{Member of CONICET, Argentina},           
           A.~T. Araudo\inst{2,3,}
           \and J. M. Paredes\inst{1}
          }

   \offprints{V. Bosch-Ramon: \\ {\em vbosch@mpi-hd.mpg.de}}
   \titlerunning{Massive protostars as gamma-ray sources}

\authorrunning{V. Bosch-Ramon et al.}  \institute{Departament d'Astronomia i Meteorologia and 
Institut de Ci\`encies del Cosmos (ICC), Universitat de Barcelona (UB/IEEC), Mart\'{\i} i Franqu\`es 1,
08028, Barcelona, Spain \and
Instituto Argentino de
Radioastronom\'{\i}a (CCT La Plata, CONICET), C.C.5, (1894) Villa Elisa, Buenos Aires,
Argentina \and Facultad de Ciencias Astron\'omicas y Geof\'{\i}sicas,
Universidad Nacional de La Plata, Paseo del Bosque, 1900 La Plata,
Argentina}

\date{Received / Accepted}


  \abstract 
{Massive protostars have associated bipolar outflows with velocities of hundreds of km~s$^{-1}$. Such outflows can produce
strong shocks when interact with the ambient medium leading to regions of non-thermal radio emission.}
{We aim at exploring under which conditions relativistic particles are 
accelerated at the terminal shocks 
of the protostellar jets and can produce significant gamma-ray emission.}  
{We estimate the conditions necessary for particle acceleration up to very high energies and 
gamma-ray production in the non-thermal hot spots of jets associated with 
massive protostars embedded in dense molecular clouds.}  
{We show that relativistic Bremsstrahlung and proton-proton collisions can make molecular clouds with massive young 
stellar objects detectable by the 
{\it Fermi}{} satellite at MeV-GeV energies and by Cherenkov telescope arrays in the GeV-TeV range.}  
{Gamma-ray astronomy can be used to probe the physical conditions in star forming regions and particle acceleration 
processes in the complex environment of massive molecular clouds.}

\maketitle
%

\section{Introduction}\label{intro}

Massive stars are formed in the dense cores of massive cold clouds (Garay \& Lizano 1999, and references therein). The
accumulation of gas in the core might proceed through previous stages of fragmentation and coalescence with the progressive
result of a massive protostar that then accretes from the environment (e.g. Bonnell et al. 1997, Stahler et al. 2000)  or
through direct accretion onto a central object of very high mass (e.g. Rodr\'\i guez et al. 2008 -RMF08-; see Shu et al. 1987
for the basic mechanism). In any case, the prestellar core is expected to have angular momentum, which would lead to the
formation of an accretion disk. The strong magnetic fields inside the cloud that thread the disk should be pulled toward the
protostar and twisted by the rotation giving rise to a magnetic tower, with the consequent outflows, as shown by numerical
simulations (e.g. Banerjee \& Pudritz 2006, 2007). 

Evidence of molecular outflows is found through methanol masers, which are likely associated with shocks formed by the
interaction with the external medium (e.g. Plambeck \& Menten 1990). However, the most important evidence for outflows comes
from the detection of thermal radio jets. These jets are observed to propagate through the cloud material along distances of
a fraction of a parsec (e.g. Mart{\'\i} et al. 1993 -MRR93-). At the end point of the jets, hot spots
due to the terminal shocks are observed in several sources. In a few cases, these hot spots are clearly non-thermal,
indicating the presence of relativistic electrons that produce synchrotron radiation (e.g. Araudo et al. 2007 -ARA07-, 
2008).

A population of relativistic electrons in the complex environment of the massive molecular cloud in which the protostar is
being formed will produce high-energy radiation through a variety of processes: inverse Compton (IC) scattering of infrared
(IR) photons from the cloud, relativistic Bremsstrahlung, and, if protons are accelerated at the shock as well, inelastic
proton-proton ($pp$) collisions. If such radiation is detectable, gamma-ray astronomy can be used to shed light on the star
forming process, the protostar environment, and cosmic ray acceleration inside molecular clouds. 

This work is devoted to discuss under what conditions the terminal shocks of jets from massive protostars can efficiently
accelerate particles, and produce gamma rays that may be detectable by the {\it Fermi} satellite and Cherenkov telescopes in
the near future. The model developed for the calculations is essentially different from the phenomelogic model presented by
ARA07, since now the dynamics of the jet termination region is characterized, the shock power estimated, the conditions for 
particle acceleration analyzed, and the role of hydrodynamical instabilities for non-thermal radiation explored. In short,
the acceleration and emission are consistently modeled together with the hydrodynamics in a more physical scenario.  Our
study is partially based on early works on non-thermal emission in young stellar objects (YSO), as those by Crusius-Watzel
(1990) and Henriksen et al. (1991), but we develop further some important aspects of the hydrodynamics-radiation relation,
and focus on massive YSOs and the feasibility of their detection with the present observational facilities. 

\section{Physical scenario}\label{sce}

A massive YSO, or a group of them, are deeply embedded into a molecular cloud. The protostar heats the cloud in such a way
that it can be detected as a strong IR source, with luminosities in the range $L_{\rm IR}\sim 10^{4-5}$~$L_{\odot}$ $\sim
10^{38-39}$~erg~s$^{-1}$, whereas the optical counterpart is obscured by the cloud. Masses and sizes of these clouds are of
the order of $\sim 10^{3}$~$M_{\odot}$ and few pc, respectively (e.g. Garay \& Lizano 1999), and the densities in the regions
in which the massive YSOs are found typically span $n_{\rm c}\sim 10^3-10^6$~cm$^{-1}$ (see Araudo et al. 2008 and references
therein).

As already mentioned, collimated outflows are present in most of massive YSOs, and thermal radiation have been detected  up
to distances of $10^{16}-10^{18}$~cm from the central star. These jets have temperatures of $\sim 10^4$~K and move at
speeds ($v_{\rm j}$) from several hundreds to $\sim 1000$~km~s$^{-1}$ (e.g. MRR93, Mart\'i, Rodr{\'\i}guez and Reipurth 1995
-MRR95-).  This shows that they are strongly supersonic flows with Mach numbers $M\ga 10$. The kinetic luminosities of these
jets are expected to be $L_{\rm j}\sim 10^{36}$~erg~s$^{-1}$ (e.g. MRR95, ARA07).

In some cases (see Araudo et al. 2008), non-thermal radio lobes have been detected
at distances of 
$Z_{\rm j}\sim$~pc, with sizes of $\sim
1$\% of this distance (MRR93, Garay et al. 2003 -GAR03-). Given the directions, sizes and distances from the core, the lobes are compatible with being produced
by the head of the jet. These radio lobes are likely generated by the strong terminal shocks of the jets, which also ionize
the shocked material. Magnetic fields should also be present, since they play an important role supporting the
cloud before the gravitational collapse, allowing the required high densities in the cores to be achieved (e.g. McKee \&
Ostriker 2007). Estimates from Zeeman measurements of the cloud magnetic field give values $B_{\rm c}\sim 0.3 n_{\rm
c5}^{1/2}$~mG (e.g. Crutcher 1999), where $n_{\rm c5}=n_{\rm c}/(10^5\,{\rm cm}^{-3})$ is the cloud density. Under these
conditions, particles can be accelerated up to relativistic energies via diffusive shock (Fermi~I) 
acceleration (DSA, e.g. Bell 1978;
see Drury 1983 for a review). These particles would produce the radiation of non-thermal nature found in the lobes, and could
generate significant emission in a broad spectral range, from radio to gamma rays. 

Some amount of thermal ultraviolet (UV)/X-ray photons is expected from the shocked material. This radiation will suffer
strong photo-electric absorption in the regions of the cloud next to the jet head and will ionize the surrounding material
improving the conditions for efficient particle acceleration (e.g. Drury et al. 1996). On the other hand, fast
radiative cooling of the shocked material can affect the lobe dynamics, and also reduce the efficiency of DSA, but could
increase the detectability of massive YSO at high energies because of the density enhancement. Finally, ionization losses of
radio emitting electrons and free-free absorption by the ionized medium could affect significantly the radio spectrum. In
some sources, free-free emission may dominate the radio band.

\begin{figure}
\includegraphics[angle=0, width=0.5\textwidth]{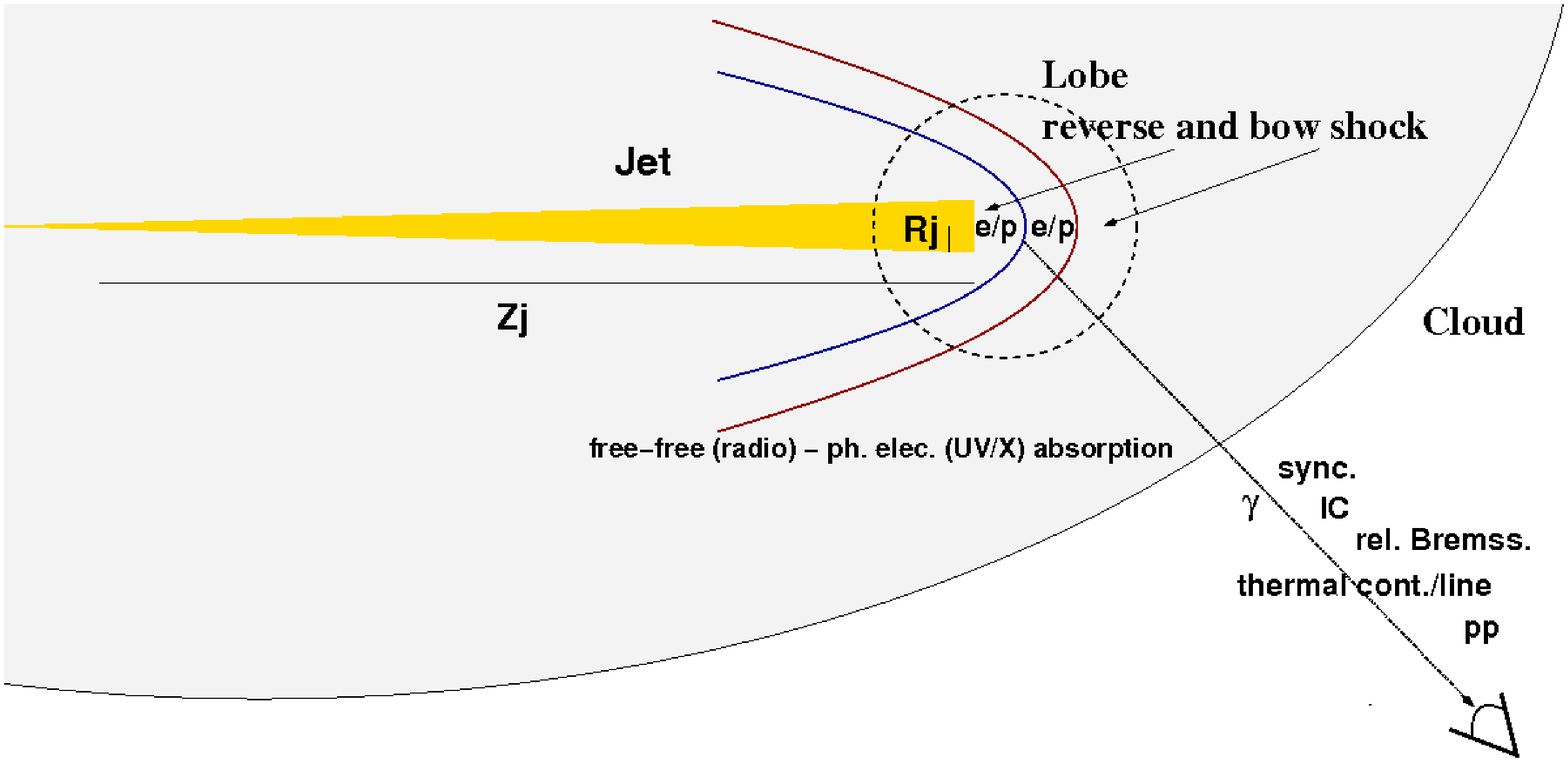}
\caption{Sketch of the termination region of the jet of a massive YSO. Two shocks of different strengths and velocities, 
depending on the jet-medium properties, will form. Electrons and protons can be accelerated in the shocks, and generate 
non-thermal emission via interaction with the ambient matter, magnetic and radiation fields. 
The shocked material will also produce thermal radiation.}\label{fig1}
\end{figure}

In Fig.~\ref{fig1} we sketch the scenario in which non-thermal emission is produced through the acceleration of
electrons and protons in the jet termination regions. 

\section{On the physical nature of the lobes}\label{phys}

We assume that the non-thermal radio lobes are the regions in which the protostellar jets terminate. The action of the jet
head on the external medium leads to two shocks, one moving in the cloud material and another one in the jet itself; these
are the bow shock and the reverse shock, respectively. These shocks would be the accelerators of the relativistic particles
generating the observed non-thermal radio emission. 

\subsection{Dynamics of the jet termination shocks}

An important parameter determining the shock characteristics is the jet (j) to cloud (c) density ratio  $\chi=n_{\rm
j}/n_{\rm c}$.  For fiducial values of the jet properties, say $v_{\rm j}\sim 10^8$~cm~s$^{-1}$,
$L_{\rm j}\sim 10^{36}$~erg~s$^{-1}$ and jet radius $R_{\rm j}\sim
10^{16}-10^{17}$~cm (assuming that lobe and jet radii are similar),  we obtain jet densities in the range $n_{\rm j}\sim
10^2-10^4$~cm$^{-3}$ at the location of the lobe; 
then, $\chi\sim 10^{-4}-10$. 

The value of $\chi$, together with $v_{\rm j}$, determines 
the speed of the
bow shock (e.g. Blondin et al. 1989):
$$
v_{\rm bs}\approx
(1+\chi^{-1/2})^{-1}v_{\rm j}\approx (0.01-0.8)v_{\rm j}
$$
\begin{equation}
~~~~~~~~~\approx 10^6-10^8\,v_{\rm j8}\,{\rm cm~s}^{-1}\,,
\end{equation}
where $v_{\rm j8}=v_{\rm j}/(10^8$~cm~s$^{-1})$. The reverse shock velocity is  
$v_{\rm r}\sim v_{\rm j}-3/4\,v_{\rm bs}$. The life time of the jet, which can be written as
$t_{\rm life}\approx Z_{\rm j}/v_{\rm bs}$, can be also expressed
as a function of $v_{\rm j8}$ and the parameters $Z_{\rm pc}=Z_{\rm j}/({\rm 1~pc})$, $\chi_{0.1}=\chi/0.1$:
\begin{equation}
t_{\rm life}\approx Z_{\rm j}/v_{\rm bs}\sim 3\times 10^{10}\,Z_{\rm pc}\,v_{\rm j8}^{-1}\,{\rm s}\,,\,\,{\rm when}\,\,\chi>1\,,
\label{lifej1}
\end{equation}
and
\begin{equation}
t_{\rm life}\approx Z_{\rm j}/v_{\rm bs}\approx 10^{11}\,Z_{\rm pc}\,v_{\rm j8}^{-1}\,\chi_{0.1}^{-1/2}\,{\rm s}\,,\,\,{\rm
when}\,\,\chi<1\,.
\label{lifej2}
\end{equation}

Equations \ref{lifej1} and \ref{lifej2} 
show that the jet advance takes place in two different regimes depending on the jet-medium density ratio, which
depends in turn on the source age.
As long as the jet lateral pressure is larger than that of the surrounding medium, the jet expands freely and thereby $n_{\rm
j}\propto R_{\rm j}^2\propto Z_{\rm j}^2$, i.e. the jet has 
a conical shape. For very young jets, when $\chi>1$, the
advancing jet head speed is constant ($v_{\rm bs}\sim v_{\rm j}$), and therefore the jet lifetime is simply 
$t\propto Z_{\rm j}$. However,
the jet head gets diluted. Since $\chi\propto Z_{\rm j}^{-2}$, at some stage $\chi<1$ and $t$ becomes
$\propto Z_{\rm j}^2$. This $Z_{\rm
j}$-$t$ dependence implies that most of the sources will be observed when $\chi<1$. 

At some point, the jet expansion is stopped by the external pressure and the jet 
density becomes roughly constant. When it happens depends on the shocked jet material pressure away from the reverse shock,
but it is expected that $\chi\ll 1$.
If values as low as $\chi\sim 10^{-4}$ are reached
the bow shock will move with a speed below the Alfven speed ($v_{\rm A}$) in the cloud and will not be a shock anymore. 

The luminosity crossing the shock surface in the jet and the cloud (reverse and bow shock, respectively) is: 
$$
L_{\rm s}=(\pi/2) R_{\rm j}^2 m_p n_{\rm j,c}v_{\rm r,bs}^3
$$
\begin{equation}
~~~~~~~~~~~~~~~~~\sim 3\times 10^{35}\,R_{16}^2\,n_{\rm j,c3}\,v_{\rm
r,bs8}^3\,{\rm erg~s}^{-1}\,,
\end{equation}
where $n_{\rm j,c3}=n_{\rm j,c}/(10^3\,{\rm cm}^{-3})$ and $v_{\rm r,bs8}=v_{\rm r,bs}/(10^8\,{\rm cm~s}^{-1})$. 
The sum of $L_{\rm s}$ from
both shocks cannot be larger than $L_{\rm j}$. 

When $\chi<1$, the reverse shock to bow shock luminosity ratio is
$\sim\chi^{-1/2}$. This fact, together with the $Z_{\rm
j}$-$t$ dependence mentioned above, 
implies that most of the time the reverse shock will
be more powerful than the bow shock. 
On the other hand, for $\chi\gg 1$ the reverse shock will be very weak and the bow shock
radiation 
very faint. Therefore, lobe detection is favored when the jet has expanded enough to reach 
$\chi\la 1$, being the reverse shock
the best place for particle acceleration unless $\chi\sim 1$, when both shocks have similar properties. 

\subsection{The postshock regions}\label{posts}

For purely adiabatic jet-medium interactions, the formation of the reverse shock is accompanied 
by a strong widening of the jet termination region as in extragalactic FRII sources (e.g. Kaiser \& Alexander 1997) 
and possibly also in microquasars (e.g. Bordas et al. 2009).
However, the conditions in massive YSOs are likely different. As noted by Blondin et al. (1989), if the cooling length 
$l_{\rm cool}$ of the
shocked material in either the reverse, the bow shock, or in both, is $<R_{\rm j}$, 
the jet head will not expand significantly. 
Using the cooling function $C(T)\sim 7\times 10^{-19}\,T^{-0.6}$ 
(e.g. Myasnikov et al. 1998) at the temperatures relevant here, 
adopting a density and speed downstream right after the shock of
$4\,n_{\rm j,c}$ and $v_{\rm r,bs}/4$, respectively 
(strong, non-relativistic and lowly magnetized shock), and assuming full ionization
in that region, $l_{\rm cool}$ is:
\begin{equation}
l_{\rm cool}\sim \frac{3\,k\,T\,v_{\rm r,bs}}{32\,n_{\rm j,c}\,C(T)}
\sim 10^{18}\,v_{\rm r,bs8}^{4.2}\,n_{\rm j,c3}^{-1}\,{\rm
cm}\,.
\end{equation}
Despite a specific-case treatment and detailed simulations would be required to characterize 
the fine evolution of the jet head, 
there is a wide range of realistic parameters for which the bow shock, and even the reverse shock, can
be radiative. This is compatible with the relatively small size of non-thermal radio lobes (see Sect.~\ref{sce}).
In case one or both shocks are not radiative, the material cools 
through adiabatic expansion farther than $R_{\rm j}$ from the shock. 
The adiabatic cooling
timescale is similar to the advection (or escape) time of the material in the downstream region:
\begin{equation}
t_{\rm esc}\sim 4\,R_{\rm j}/v_{\rm r,bs}=4\times 10^8\,R_{\rm j16}v_{\rm r,bs8}^{-1}\,{\rm s}\,.
\label{tesc}
\end{equation}

Because of radiative cooling, the compression ratio $\Lambda$, 
or downstream to upstream density ratio, will grow with the distance 
downstream from $\Lambda=4$ to $\sim 3\times 10^3\, v_{\rm
s8}^2$, or to $\sim 20\,v_{\rm s8}\,n_{\rm
j,c3}^{1/2}\,B_{-3}^{-1}$. The former upper limit for $\Lambda$ comes from the fact that the temperature stops falling 
around $T\sim 10^4$~K, and the latter from the enhancement of the magnetic field ($B=10^{-3}\,B_{-3}$~G) pressure, 
which limits the compression (see Blondin et al. 1989).

When $\chi<1$, the material downstream is $1/\chi$ to $\Lambda_{\rm bs}/\chi\Lambda_{\rm r}$ times denser in the bow shock
than in the reverse shock. 
Since a force is exerted by the downstream material of the reverse shock 
on that of the bow shock,  
Rayleigh-Taylor (RT) instabilities in the contact discontinuity between both shocks 
can develop. This phenomenon will distort the jet head on a timescale: 
\begin{equation}
t_{\rm RT}\sim R_{\rm j}/\chi'^{1/2}\,v_{\rm r}\sim 
3\times 10^9\,R_{\rm j16}\,\chi_{0.001}'^{-1/2}\,v_{\rm j8}^{-1}\,{\rm s}\,,
\end{equation}
where $\chi_{0.001}'=(\chi\Lambda_{\rm r}/\Lambda_{\rm bs})/0.001$ is the effective density jump 
in the contact discontinuity. 
Therefore,
for $t_{\rm life}>t_{\rm RT}$, complex shock structures should form (Blondin et al. 1989) accompanied by
strong mixing of material from both shocks. 
This mixing can lead to an increase of the effective density downstream the reverse shock. 
To account for this, we have simply
parameterized the density downstream the reverse shock 
as $4\,n_{\rm j}<4\,F\,n_{\rm j}<\Lambda_{\rm bs}\,n_{\rm c}$, where $F$ is a free parameter that says how much the 
density departs from the adiabatic value, $4\,n_{\rm j}$.
Downstream the
bow shock, mixing should not affect significantly the density, and hence $F\sim 1$.

Concerning the dynamic role of the jet and cloud magnetic fields at the jet termination region, 
$B_{\rm j}$ and $B_{\rm c}$, respectively, we will assume
hereafter that they are negligible. This applies as long as $B_{\rm j,c}\ll B_{\rm eq}$, where: 
\begin{equation}
B_{\rm eq}=30\,L_{\rm j36}^{1/2}\,R_{\rm
j16}^{-1}\,v_{\rm j8}^{-1/2}\,{\rm mG} 
\label{equip}
\end{equation}
is the magnetic field of equipartition between the magnetic and the 
jet kinetic energy density. If $B_{\rm j,c}\ll B_{\rm eq}$, it is also 
expected that the magnetic field in the shock regions, $B$, should be well below 
$B_{\rm eq}$\footnote{Note that $B_{\rm eq}$ relates
to the jet kinetic energy density, not to the one of the (unshocked) cloud. 
In the latter, the magnetic field may be indeed dynamically more important.}.

\section{Particle acceleration in the lobes}\label{mod}

\subsection{Acceleration and cooling processes}\label{acc}

The energies and luminosities that non-thermal particles may achieve depend on the efficiency of particle acceleration, which
depends in turn on the lobe properties. 
As noted above, DSA can operate in the fast strong shocks at the termination of the jets
accelerating particles up to relativistic energies. 
For a parallel, lowly magnetized (i.e. $v_{\rm r,bs}\gg v_{\rm A}$ upstream), non-relativistic strong shock, in the test 
particle approximation and with diffusion coefficient $D$, the acceleration rate is (e.g. Protheroe 1999):
$$
\dot{E}_{e,p\;\rm gain}=\frac{3}{20}\,d^{-1}\,\left(\frac{v_{\rm r,bs}}{c}\right)^2\,q\,B\,c
$$
\begin{equation}
~~~~~~~~~~~~\approx1.5\times 10^{-5}\,d^{-1}\,v_{\rm r,bs8}^2\,B_{-3}\,{\rm GeV~s}^{-1}\,,
\end{equation}
where $e$ and $p$ stand for electrons and protons, respectively, and 
$d=D/D_{\rm B}$, with $D_{\rm B}=c r_{\rm g}/3$ being the diffusion coefficient in the Bohm limit 
($r_{\rm g}=E/qB$ is the gyroradius of a particle with energy $E$). 
The acceleration timescale can be written as:
\begin{equation}
t_{\rm gain}=E/\dot{E}_{e,p\;\rm gain}\approx 6.7\times 10^4\,d\,v_{\rm r,bs8}^{-2}\,B_{-3}^{-1}E_{\rm GeV}\,\,{\rm s}\,,
\end{equation}
where $E_{\rm GeV}=E/(1\,{\rm GeV})$.

At the maximum energy of particles, $t_{\rm gain}$ 
becomes equal to the shortest timescales among
synchrotron, IC (Thomson regime) and relativistic Bremsstrahlung losses for electrons
(see Blumenthal \& Gould 1970), 
inelastic $pp$ collision losses 
for protons (see Kelner et al. 2006), and diffusive particle escape and jet life time for both electrons and protons. 
The relevant loss timescales, $t_{\rm loss}=-E/\dot{E}_{\rm loss}$, 
are given by the following expressions:
\begin{equation}
t_{\rm sync}\approx 4\times 10^{11}\,B_{-3}^{-2}\,E_{\rm GeV}^{-1}\,{\rm s}\,,
\end{equation}
\begin{equation}
t_{\rm IC}\approx 1.6\times 10^{13}\,u_{\rm IR-9}^{-1}\,E_{\rm GeV}^{-1}\,{\rm s}\,,
\end{equation}
\begin{equation}
t_{\rm Brems}\approx 3.5\times 10^{10}\,n_{\rm j,c3}^{-1}\,F_{10}^{-1}\,{\rm s}\,,
\end{equation}
\begin{equation}
t_{pp}\approx 5\times 10^{10}\,n_{\rm j,c3}^{-1}\,F_{10}^{-1}\,{\rm s}\,,
\end{equation}
\begin{equation}
t_{\rm diff}\approx 1.5\times 10^{12}\,d^{-1}\,R_{\rm j16}^2\,B_{-3}\,E_{\rm GeV}^{-1}\,{\rm s}\,,
\label{acc1}
\end{equation}
\begin{equation}
t_{\rm life}\approx 10^{11}\,Z_{\rm pc}\,v_{\rm bs8}^{-1}\,{\rm s}\,,
\label{acc1}
\end{equation}
where $u_{\rm IR-9}=u_{\rm IR}/10^{-9}\,{\rm erg~cm}^{-3}$ is the 
energy density of the IR photon field and $F_{10}=F/10$ 
(recall $F_{10}$=0.1 in the bow shock). Concerning the densities, 
the compression of the material due to the shock is already accounted.
Under the adopted assumptions, particles spend on average $\sim 4$ times longer in the downstream than in the upstream
region, and the conditions downstream the shock will determine the particle evolution. 
Adiabatic cooling is not considered in
the accelerating region. 

From the timescales presented above, we obtain the following maximum energies 
depending on the dominant energy loss mechanism and source age:
\begin{equation}
E_{\rm max~sync}\approx 2.4\times 10^3\,d^{-1/2}\,v_{\rm r,bs8}\,B_{-3}^{-1/2}\,{\rm GeV}\,,
\end{equation}
\begin{equation}
E_{\rm max~IC}\approx 1.5\times 10^4\,d^{-1/2}\,v_{\rm r,bs8}\,u_{\rm IR-9}^{-1/2}\,B_{-3}^{1/2}\,{\rm GeV}\,,
\end{equation}
\begin{equation}
E_{\rm max~Brems}\approx 5.2\times 10^5\,d^{-1}\,v_{\rm r,bs8}^2\,B_{-3}\,n_{\rm j/c3}^{-1}\,F_{10}^{-1}\,{\rm GeV}\,,
\end{equation}
\begin{equation}
E_{\rm max~pp}\approx 7.5\times 10^5\,d^{-1}\,v_{\rm r,bs8}^2\,B_{-3}\,n_{\rm j/c3}^{-1}\,F_{10}^{-1}\,{\rm GeV}\,,
\end{equation}
\begin{equation}
E_{\rm max~diff}\approx 4.7\times 10^3\,d^{-1}\,v_{\rm r,bs8}\,B_{-3}\,R_{\rm j16}\,{\rm GeV}\,,
\label{acc2}
\end{equation}
\begin{equation}
E_{\rm max~life}\approx 1.5\times 10^5\,d^{-1}\,Z_{\rm pc}\,v_{\rm r,bs8}\,B_{-3}\,{\rm GeV}\,,
\label{acc2}
\end{equation}
This 
shows that, if $d\sim 1$, electrons and protons can reach very high energies. It turns out that 
synchrotron and diffusive escape are the dominant mechanisms to limit acceleration, although relativistic Bremsstrahlung 
and $pp$ collisions can dominate for large densities. 
We notice that in radiative shocks, $R_{\rm j}$ should
be substituted by $l_{\rm cool}$ in Eqs.~(\ref{acc1}) and (\ref{acc2}).

We note that the accelerated proton to electron number ratio $a$ cannot be established from first principles. It is
considered here as a phenomenologic parameter.

\subsection{Required conditions for efficient DSA}

In DSA particles are scattered by magnetic inhomogeneities back to the shock before escaping downstream/upstream. These
inhomogeneities should not move faster than the shock itself, since otherwise the particles would not be affected by the
shock velocity jump. If it were the case, stochastic Fermi~II particle acceleration could be at work (Fermi 1949). However,
in that case most of the jet kinetic energy will not be available and little energy will go to non-thermal particles. 
Therefore,
efficient particle acceleration requires the shock to be  
super-Alfvenic in the upstream region, i.e. $v_{\rm A}\ll v_{\rm r,bs}$, where: 
\begin{equation}
v_{\rm A}\approx \sqrt{\frac{B^2}{4\pi X_{\rm i} n_{\rm j,c} m_{\rm p}}}\approx 2.1\times 10^7 B_{-3} 
X_{\rm i0.1}^{-1/2} n_{\rm j,c3}^{-1/2}\,{\rm
cm~s}^{-1}\,, 
\end{equation}
and $X_{\rm i}=0.1\,X_{\rm i0.1}$ is the ratio of ionized to total density. 
Thus, $X_{\rm i}$ should not be too low. 

Downstream the shock, the material is likely ionized through particle collisions, and  the magnetic field is expected to be
quite disordered. 
Upstream the shock, UV/X-ray radiation produced in the shock region can photo-ionize the medium unless 
$v_{\rm bs}\la 10^7$~cm~s$^{-1}$ (or $\chi\la 0.01\,v_{\rm
j8}^{-2}$), since then photons cannot ionize hydrogen. 
From the recombination and photo-ionization timescales, $t_{\rm re}\sim 10^{11}\,X_{\rm i0.1}^{-1}\,n_{\rm j,c3}^{-1}$~s
and 
$t_{\rm ph}\sim 10^7\,R_{\rm 16}^{2}\,v_{\rm s8}^{2}\,L_{35}^{-1}$~s, respectively
(the recombination rate and ionization cross section are 
given in Seaton 1959 and Morrison \& McCammon 1983; $L_{35}$ corresponds here to the ionizing photon field luminosity), it seems likely
that upstream the shock 
$t_{\rm ph}<t_{\rm re}$, and therefore $X_{\rm i}\rightarrow 1$. In the radiative regions downstream the shock, 
$X_{\rm i}$ may become much smaller than 1.

As noted, the magnetic field should have some level of inhomogeneity: the scattering centers that isotropise particles in
both sides of the shock. 
Magnetic inhomogeneities can develop upstream due to relativistic
particle streaming (e.g. Lucek \& Bell 2000), but they should not be suppressed by wave damping (see Reville et al. 2007).
These inhomogeneities would be advected downstream the shock, thus they would be also present there.

For very high densities, Coulombian/ionization energy losses should not suppress acceleration 
from suprathermal energies (e.g. Drury et al. 1996). For that, the following condition should be provided:
\begin{equation} 
t_{\rm ion}\sim
3\times 10^{11}\,(m_{p,e}/m_{e})\,E_{\rm GeV}\,n_{\rm j,c3}^{-1}\,F_{10}^{-1}\,{\rm s}\,>t_{\rm gain}\,,
\end{equation}
which implies: 
\begin{equation}
n_{\rm j,c}< 4\times 10^9\,(m_{p,e}/m_{e})\,F_{10}^{-1}\,v_{\rm s8}^2\,B_{-3}\,{\rm cm}^{-3}\,.
\end{equation}

The detection of radio emission from electrons with energy:
\begin{equation}
E\approx 0.6\,\nu_{\rm 5~GHz}^{1/2}\,B_{-3}^{-1/2}\,{\rm GeV}\,\gg 0.511\,{\rm MeV}\,, 
\label{tipf}
\end{equation}
where $\nu$ is the radiation frequency ($\nu_{\rm
5~GHz}=\nu/(5\,{\rm GHz})$), is evidence for efficient particle acceleration and hence some
degree of ionization and $B$-inhomogeneity at least in some sources. However, $d$, which relates to the 
$B$-inhomogeneity, may be
small or even energy dependent, not allowing particle acceleration to be efficient beyond 
energies at which electrons produce synchrotron radio emission. 
Also, if $l_{\rm
cool}\ll R_{\rm j}$ or mixing were very efficient (large $F$), electron acceleration could also stop at radio emitting
energies because of large densities and strong relativistic Bremsstrahlung losses. In such a case, protons could not reach 
very high energies neither due to strong $pp$ cooling. 

\section{Non-thermal emission in the lobes}\label{nont}

\subsection{The fate of accelerated particles}\label{nont}

The electrons and protons accelerated by DSA have at injection an energy spectrum $Q(E)\propto E^{-\Gamma}$ up to $E_{\rm
max}$, with $\Gamma\sim 2$ (e.g. Drury 1983) and total luminosity $L_{\rm nt}^{e,p}<L_{\rm s}$. They mainly accumulate downstream the shock,
in a region that here will be considered homogeneous and with a typical size $\sim R_{\rm j}$. When protons are present,
secondary electrons and positrons ($e^\pm$) are injected with almost a power-law in energy in the region in which these
protons interact significantly. The injection luminosity and the maximum effective energy of $e^\pm$ will be about a
half of the luminosity going to $\pi^0$-decay gamma-rays and $\sim 0.1$ the energy of protons, respectively (see Kelner et
al. 2006). All these particles evolve under the downstream magnetic, matter and radiation fields, losing energy in the form
of synchrotron radiation, relativistic Bremsstrahlung and IC emission in the case of electrons, and high-energy photons,
neutrinos, $e^\pm$ and other secondary particles via $pp$ collisions in the case of protons. 

Only the radiation from the region closer than $R_{\rm j}$ from the shock will be computed. At distances larger than
$R_{\rm j}$, particles cool through adiabatic losses due to the re-expansion of the shocked material, producing scarce
emission.  If densities are large enough (i.e. strongly radiative shocks, $F\gg 1$), electrons will cool fast via
ionization/Coulombian losses and relativistic Bremsstrahlung, and protons through $pp$ collisions, and they will not reach
the adiabatic cooling region far downstream. 

Far upstream the bow shock, particles with $t_{\rm diff}$ shorter than the dominant cooling timescale $t_{\rm loss}$ could
escape into the cloud. Some of the highest energy electrons and protons would escape from the accelerator in this way. 

The particle distribution, 
$N(E,t)$, can be obtained solving the transport equation (e.g. Ginzburg \& Syrovatskii 1964):
\begin{equation}
\frac{\partial N\left(t,E\right)}{\partial t}+\frac{\partial \left[b(E) 
N\left(t,E\right)\right]}{\partial E}+\frac{N\left(t,E\right)}{t_{\rm esc}}=Q(E)\,,
\label{Ginz}
\end{equation}
where $Q(E)$ is assumed to be constant in time
and $b(E)$ includes all the cooling rates $-E/t_{\rm loss}$ relevant for $N(E,t)$, i.e. 
synchrotron, relativistic Bremsstrahlung and IC processes for electrons and $pp$ collisions for protons. 
The adiabatic cooling,
which would operate far from the shock as described above, is not considered here.
The parameter $t_{\rm esc}$, the escape time,
is the advection timescale (see Eq.~\ref{tesc}), different from $t_{\rm diff}$, which was used to compute $E_{\rm max~diff}$. 
This escape time is the dominant timescale for the removal of particles from the emitting region. 
Since in general $t_{\rm esc}\ll t_{\rm life}$, particles will be in the steady state, i.e. $\partial
N\left(t,E\right)/\partial t=0$.

\subsection{Radiation luminosities and spectral energy distributions}\label{lum}

\subsubsection{Luminosities}

Adopting an efficiency $f_{\rm nt}^{e,p}=0.1\,f_{\rm nt0.1}^{e,p}$ for transferring shock luminosity to non-thermal particles 
(electrons or protons), 
where $f_{\rm nt}^{e,p}<1$, we get:
\begin{equation}
L_{\rm nt}^{e,p}=f_{\rm nt}^{e,p}\,L_{\rm s}=10^{35}\,f_{\rm nt0.1}^{e,p}\,L_{\rm s36}\,{\rm erg~s}^{-1}\,,
\end{equation}
where $L_{\rm s36}=L_{\rm s}/(10^{36}\,{\rm erg}\,{\rm s}^{-1})$.
Since particle escape has the dominant dynamical timescale, we can
roughly calculate the relativistic Bremsstrahlung
and $pp$ photon bolometric luminosities in the lobe:  
\begin{equation}
L_{\rm Brems}\sim 10^{33}\,f_{\rm nt0.1}^e\,F_{10}\,R_{\rm j16}\,v_{\rm
s8}^{-1}\,n_{\rm j,c3}\,L_{\rm s36}\,{\rm erg~s}^{-1}\,, 
\label{lbr}
\end{equation}
\begin{equation}
L_{pp}\sim 10^{32}\,f_{\rm nt0.1}^p\,F_{10}\,R_{\rm j16}\,v_{\rm s8}^{-1}\,n_{\rm j,c3}\,L_{\rm s36}\,{\rm
erg~s}^{-1}\,.
\label{lpp}
\end{equation} 
These luminosities cannot be higher than $L_{\rm nt}^e$ and $\sim 1/3\,L_{\rm nt}^p$ for relativistic Bremsstrahlung 
and $pp$ collisions, respectively.
These two mechanisms contribute mainly to the high-energy part of the spectrum.
Regarding the synchrotron/IC luminosities, 
in general, when 
\begin{equation}
r(E_{\rm max})=t_{\rm sync,IC}/{\rm min}[t_{\rm esc},t_{\rm Brems},t_{\rm IC,sync}]<1\,:
\end{equation}
\begin{equation}
L_{\rm sync,IC}\sim 10^{35}\,f_{\rm nt0.1}^e\,L_{\rm s36}\,{\rm erg~s}^{-1}\,.
\label{lsic1}
\end{equation}
Otherwise:
\begin{equation}
L_{\rm sync,IC}\sim 10^{35}\,f_{\rm nt0.1}^e\,r^{-1}\,L_{\rm s36}\,{\rm erg~s}^{-1}\,.
\label{lsic2}
\end{equation}
The IC luminosities 
will be a minor component unless $u_{-9}\gg 1$. Synchrotron and IC contribute to the low- and the high-energy parts of the
spectrum, respectively.
Note that $f_{\rm nt}^e$ and $f_{\rm nt}^p$ 
may actually be very different in some sources (as inferred from the proton-to-electron number ratio 
in cosmic rays, i.e. $a\approx 100$, Ginzburg \& Syrovatskii 1964).

The highest energy electrons and protons may escape from the lobe and radiate in the cloud, 
although the corresponding luminosities depend on the escape
probability, which is difficult to quantify. 

\subsubsection{Spectral energy distributions}

If synchrotron or IC losses dominate at $E<E_{\rm max}$ for electrons, there is a break in the particle energy distribution
$N(E,t)$ at $E\sim E_{\rm b}$, in which $t_{\rm sync,IC}$ becomes the shortest timescale. Above $E_{\rm b}$,
$N(E,t)\propto E^{-(\Gamma+1)}$, which yields a spectral energy distribution (SED) of the radiation that is
$\propto\epsilon^{(2-\Gamma)/2}$ for synchrotron and IC dominance, and $\propto \epsilon^{1-\Gamma}$ for relativistic
Bremsstrahlung  ($\epsilon$ is the photon energy). 

For electron energies $E<E_{\rm b}$, and at any energy for protons,
advection, and relativistic Bremsstrahlung or $pp$ collisions, lead to $N(E,t)\propto E^{-\Gamma}$, which yields a
SED $\propto \epsilon^{(3-\Gamma)/2}$ (synchrotron/IC), and $\propto \epsilon^{2-\Gamma}$ (both relativistic Bremsstrahlung and 
$pp$
collisions). Below $\epsilon\sim m_{e}c^2\sim 0.5$~MeV and $\epsilon\sim m_{\pi}c^2\sim 140$~MeV, the SEDs of relativistic
Bremsstrahlung and $pp$ collisions become roughly $\propto \epsilon$ and $\propto\epsilon^2$, respectively. 

Dominant
ionization/Coulombian losses, relevant only for electrons in our context, lead to $N(E,t)\propto E^{1-\Gamma}$,
yielding a SED $\propto \epsilon^{2-\Gamma/2}$ and $\propto \epsilon^{3-\Gamma}$ for synchrotron/IC and relativistic
Bremsstrahlung, respectively. 

\subsubsection{Requirements from observed spectra}

The non-thermal nature of the observed radio spectra in several specific cases implies that strong free-free
absorption of radio emission should not occur in the lobe or in the surroundings. Far from the lobe 
the ionization degree should be small. Close to the lobe, the medium is ionized and
it is necessary to account for free-free
absorption, which is expected to be large downstream the bow shock. The
free-free opacity there can be written as (Rybicki \& Lightman 1979): 
\begin{equation}
\tau_{\rm ff}\sim 0.03\,T_5^{-1.5}\,\nu_{\rm 5~GHz}^{-2}\,X_{\rm i0.1}^2\,n_{\rm c5}^2\,\Lambda_{10}^2\,l_{16}\,,
\end{equation} 
where
$T_5=T/10^5$~K, and $l_{16}=l/(10^{16}\,{\rm cm})$ is the typical size of the region. Also, suppression of emission 
at frequencies $\nu<2\times 10^8\,n_{\rm j,c5}X_{\rm i0.1}/B_{-3}$~Hz, due to the Tsytovich-Razin effect, should be
considered.

Another condition that should be fulfilled given the observed radio spectra is 
that either electron escape or relativistic
Bremsstrahlung should dominate over Coulombian/ionization losses at low electron energies, i.e. 
$\alpha\sim 0.5$ ($F_{\nu}\propto \nu^{-\alpha}$). From the timescale ratios: 
\begin{equation}
t_{\rm ion}/t_{\rm esc}\sim 8\times 10^2\,E_{\rm GeV}\,n_{\rm j,c3}^{-1}\,F_{10}^{-1}\,v_{\rm r,bs8}^{-1}\,\,{\rm and}\,\,
\end{equation}
\begin{equation}
t_{\rm ion}/t_{\rm Brems}\sim 9\,E_{\rm GeV}\,,
\end{equation}
it is seen that upstream and downstream the shock, radio emitting electrons are dominated by escape or 
relativistic Bremsstrahlung losses. Note that sources with spectra harder than 
$F_{\nu}\propto \nu^{-0.5}$ may be still explained in a non-thermal scenario 
by moderate free-free absorption and/or ionization cooling (or a thermal component, see below). 

\subsection{Deriving the magnetic field strength}

Assuming a value for $L_{\rm  nt}^e$, plus some additional simplifying assumptions, a formula has been obtained 
to derive the magnetic field strength consistent with the observed radio fluxes.
Taking the radio fluxes at a certain frequency,
a particle energy distribution with $\Gamma\sim 2$ and normalized with the total energy
$\sim L_{\rm nt}^e\,t_{\rm esc}$, the synchrotron
power for one electron ($\dot{E}\approx 4.1\times 10^{-15}\,B_{-3}^2\,E_{\rm GeV}^2$~erg~s$^{-1}$), and the reasonable
simplification that electrons of energy $E$ produce only photons of frequency $\nu\approx 5\times 10^9\,B_{-3}\,E_{\rm
GeV}^2$~Hz, we obtain:
\begin{equation}
B\sim 0.04\,(L_{\rm nt35}^e)^{-2/3}\,R_{\rm j16}^{-2/3}\,v_{\rm s8}^{2/3}\,\nu_{\rm 5~GHz}^{1/3}\,d_{\rm 3~kpc}^{4/3}\,F_{\rm 
\nu~mJy}^{2/3}\,{\rm mG}\,,
\label{mag}
\end{equation}
where $d_{\rm kpc}$ is the distance the source in kpc, and $F_{\rm \nu~mJy}$ the flux in mJy at the relevant 
frequency. In case the radio emitting leptons are secondary $e^{\pm}$, then in Eq.~(\ref{mag}) it should be changed:
\begin{equation}
L_{\rm nt}^e\rightarrow 0.1\,(t_{\rm esc}/t_{pp})\,L_{\rm nt}^p\,.
\end{equation}

\section{Thermal emission in the lobes}

The shocked material is heated up to $T\approx 2.3\times 10^7\,v_{\rm s8}^2$~K ($2.4\,v_{\rm s8}^2$~keV) and generates line
and thermal Bremsstrahlung continuum emission. The total luminosity cannot overcome $L_{\rm s}$, and it might be much smaller
if the reverse shock is adiabatic and the bow shock relatively slow. This radiation can be absorbed in the cloud core, since
the photo-electric opacity coefficient is $\tau_{\rm X}\sim 20(\epsilon_{\rm keV})^{-2.5}N_{\rm H23}$ (within a factor of 2
in the relevant energy range; see Morrison \& McCammon 1983), where $\epsilon_{\rm keV}=\epsilon/{\rm keV}$ is the photon
energy and $N_{\rm H}=10^{23}\,N_{\rm H23}=10^{23}\,n_{\rm c5}\,l_{\rm 18}$~cm$^{-2}$ the neutral hydrogen column density of
the lobe surroundings. 

The density in the shock regions should neither be too high, to avoid suppression of the acceleration due to strong cooling,
nor too low, to avoid that $v_{\rm A}\ga v_{\rm r,bs}$. Actually, there is room for the 
shocks to be radiative, producing thermal X-rays that may
escape the cloud, without necessarily suppressing particle acceleration.

The fact that thermal radio emission should not overcome the non-thermal component at the same frequencies put some
constraints on the scenario. The thermal SED peaks at energies up to $\sim$~keV and is harder than the synchrotron SED,
which  should generally peak at lower energies. This means that the total $L_{\rm sync}$ must be $\ll L_{\rm s}$ ($L_{\rm
sync}\la 10^{-4}\,L_{\rm s}$ for $\Gamma=2$), if the non-thermal radio emission is to be dominated by the thermal one. This
condition is hard to fulfill unless $f_{\rm nt}^e\ll 1$. It is worth noting that a radio spectra harder than $\nu^{-0.5}$
may be non-thermal radiation  contaminated by a thermal component. 

As mentioned in Sect.~\ref{intro}, several massive YSOs present non-thermal radio emission. Two among the most relevant of
them are studied in next section: IRAS~16547$-$4247 and the complex source HH~80$-$81.

\section{Application to IRAS~16547$-$4247 North and HH~80}\label{gamma}

The model described in Sect.~\ref{nont} is applied now to the northern radio lobe of the massive YSO  IRAS~16547$-$4247
(IRAS-N), and to the radio lobe HH~80 in the complex source HH~80$-$81. Both lobes have a clear non-thermal nature (e.g.
GAR03; MRR93). IRAS-N shows also an extended structure pointing to the South-East, and HH~80 has a similar non-thermal source
very nearby, HH~81. We will not consider here either the South-East extension of IRAS-N nor HH~81. Note however that the
extension in IRAS-N may be in fact a fore/background object (RMF08), and HH~81 could be part of the expected complex structure
of the jet termination region (see Sect~\ref{phys}; see also Heathcote et al. 1998 -HRR98- for a detailed optical
study of the HH~80$-$81 complex).

\subsection{IRAS-N and HH~80 properties}

We focus on IRAS-N and HH~80 because both sources are at the two extremes of the density parameter range presented above.
IRAS-N is embedded in a very dense cloud, with $n_{\rm c}\sim 5\times 10^5$~cm$^{-3}$ (GAR03), whereas HH~80 is thought to be
close to the border of a cloud, in a more diluted medium with $n_{\rm c}\sim 10^2-10^3$~cm$^{-3}$ (e.g. MRR93; HRR98;  Pravdo
et al. 2004 -PTM04-). The distances to IRAS-N and HH~80 are $d\sim 2.9$~kpc and $\sim 1.7$~kpc (RMF08; MRR93), and the central
stars show luminosities of $L_*\approx 5\times 10^{38}$ and $8\times 10^{37}$~erg~s$^{-1}$, respectively (GAR03; MRR93). This
radiation provides the main contribution to the infrared emission in the lobes, yielding photon energy densities there of
$u\sim 2\times 10^{-9}$~erg~cm$^{-3}$ for IRAS-N, and $2\times 10^{-12}$~erg~cm$^{-3}$ for HH~80. The distances from the
central star to the lobes are $Z_{\rm j}\approx 5\times 10^{17}$~cm for IRAS-N, accounting for a jet inclination angle of
84$^\circ$ (Garay et al. 2007), and $\sim 10^{19}$~cm for HH~80, with a not so well constrained jet inclination (see however
HRR98). The lobe sizes for IRAS-N and HH~80 are about $R_{\rm j}\approx 1.6\times 10^{16}$~cm and $5\times 10^{16}$~cm,
(RMF08; MRR93), the velocities of the jets are expected to be around $v_{\rm j}\sim 5\times 10^7$~cm~s$^{-1}$
and $\sim 10^8$~cm~s$^{-1}$, and the velocities of the bow shocks would be $v_{\rm bs}<10^7$~cm~s$^{-1}$ and $\sim 5\times
10^7$~cm~s$^{-1}$, respectively (RMF08; MRR93, MRR95, HRR98). IRAS-N has not been detected in X-rays (see ARA07), whereas HH~80
has been detected by {\it XMM} (PTM04).

\subsubsection{Derived parameters}

The values of $v_{\rm bs}$ for IRAS-N imply that $\chi\la 0.04$ and $v_{\rm r}\sim v_{\rm j}$. In fact, from the inferred age of
IRAS-N, $t_{\rm life}\sim 10^{11}$~s (Garay et al. 2007), $Z_{\rm j}$ and $v_{\rm j}$, a particle density of  $n_{\rm j}\sim
5\times 10^3$~cm$^{-3}$ can be derived, i.e. $\chi\sim 0.01$ and $v_{\rm bs}\sim 5\times 10^6$~cm~s$^{-1}$, consistent with
the limit given above. Such a value for  $n_{\rm j}$, together with $R_{\rm j}$ and $v_{\rm j}$, renders a $L_{\rm j}\sim
5\times 10^{35}$~erg~s$^{-1}$ for IRAS-N. Given that $v_{\rm r}\sim v_{\rm j}$ in this source, it is the case that $L_{\rm
s}\sim L_{\rm j}$ in the reverse shock. The bow-shock luminosity will be $\sim 4\times 10^{34}$~erg~s$^{-1}$.

For HH~80, since $n_{\rm c}\sim n_{\rm j}$, then $\chi\sim 1$, and therefore we have that $v_{\rm r}\sim v_{\rm bs}$. In the
case of HH~80 this means that both the reverse shock and the bow shock may contribute to the non-thermal radiation (see also
HRR98). Accounting for $\chi$ and $v_{\rm j}$, it can be inferred  $t_{\rm life}\sim 3\times 10^{11}$~s, not far from the
value discussed in MRR93; also, $v_{\rm r}\sim v_{\rm bs}$. Taking $L_{\rm j}\sim 2\times 10^{36}$~erg~s$^{-1}$ (MRR95), we
get $n_{\rm j}\sim 4\times 10^2$~cm$^{-3}$ and, since $\chi\sim 1$, $n_{\rm c}\sim 4\times 10^2$~cm$^{-3}$. This latter value
is between those inferred from X-ray and optical observations (see the discussion in PTM04). We will treat the bow shock and
the reverse shock in HH~80 as a single physical system, with speed $5\times 10^7$~cm~s$^{-1}$ and $L_{\rm s}=L_{\rm j}$. The
full list of the relevant properties of IRAS-N and HH~80, together with the derived parameters, is presented in
Table~\ref{tab0}.

Interestingly, the central star is brighter in IRAS-N than in HH~80, but $L_{\rm j}$ seems smaller in the former. This
could be related to the larger density of the environment in IRAS-N. This may have induced jet deceleration through, e.g.,
medium mass entrainment in the jet. Nevertheless, the uncertainties are large and no strong conclusions can be derived in
this regard.

It is worth mentioning that very powerful, slow molecular outflows with luminosities
10--100 times larger than $L_{\rm j}$ have  
been detected in the two sources (see RMF08 and references therein).  

 \begin{table}[]
  \begin{center}
  \caption[]{IRAS-N and HH~80 properties and derived parameters (see the text for details)}
  \label{tab0}
  \begin{tabular}{lll}
  \hline\noalign{\smallskip}
  \hline\noalign{\smallskip}
  & IRAS-N    & HH~80    \\
  \hline\noalign{\smallskip}
$n_{\rm c}$ [cm$^{-3}$] & $5\times 10^5$ & $4\times 10^2$ \\
$d$ [kpc] & 2.9 & 1.7 \\  
$L_*$ [erg~s$^{-1}$] & $5\times 10^{38}$ & $8\times 10^{37}$ \\
$u$ [erg~cm$^{-3}$] & $2\times 10^{-9}$ & $2\times 10^{-12}$ \\
$Z_{\rm j}$ [cm] & $5\times 10^{17}$ & $10^{19}$ \\
$n_{\rm j}$ [cm$^{-3}$] & $5\times 10^5$ & $4\times 10^2$ \\
$R_{\rm j}$ [cm] & $1.6\times 10^{16}$ & $5\times 10^{16}$\\
$v_{\rm j}$ [cm~s$^{-1}$] & $5\times 10^7$ & $10^8$ \\
$v_{\rm bs}$ [cm~s$^{-1}$] & $5\times 10^6$ & $5\times 10^7$ \\
$v_{\rm r}$ [cm~s$^{-1}$] & $5\times 10^7$ & $5\times 10^7$ \\
$t_{\rm life}$ [s] & $10^{11}$ & $3\times 10^{11}$ \\
$n_{\rm j}$ [cm$^{-3}$] & $5\times 10^3$ & $4\times 10^2$ \\
$\chi$   & 0.01 & 1 \\
$L_{\rm j}$ [erg~s$^{-1}$] & $5\times 10^{35}$ &$2\times 10^{36}$ \\
  \noalign{\smallskip}\hline
  \end{tabular}
  \end{center}
\end{table}

\subsection{Estimates of the emission in IRAS-N and HH~80}\label{est}

\subsubsection{Constraints on the non-thermal population}

In order to model the radio emission from IRAS-N and HH~80, and to compute the radiation at high and very high energies, the
values of $B$ and $L_{\rm nt}^{e,p}$ for both sources are required. $B$ should be well below $B_{\rm eq}$ (see Eq.~\ref{equip}).
The non-thermal luminosity $L_{\rm nt}^{e,p}$ will be taken equal to 
$0.1\,L_{\rm s}$, or $f_{\rm nt0.1}^{e,p}=1$. The observed radio
fluxes and spectral indices are $\sim 8.7$ and 3~mJy at 8~GHz and $\alpha\sim 0.5$ and 0.3, respectively (GAR03; MRR93). The
value of the index $\Gamma$ for the radio emitting particles can be obtained from $\alpha$, as shown in Sect.~\ref{nont}.  Fixing
$L_{\rm nt}^e$ and knowing the radio fluxes, $B$ can be estimated to zeroth order with Eq.~(\ref{mag}). We note that
equipartition with the relativistic particles would lead to magnetic fields of the order of 0.1-1~mG in the emitting regions
(see ARA07 for the case of IRAS-N).

The nature of the radio emitting particles may be primary electrons accelerated in the shock, or secondary $e^\pm$ from
$pp$ collisions. If primary electrons dominate the production of the radio emission, the relativistic proton population will
be constrained by the fact that secondary radiation cannot overcome that of primary electrons (although $pp$ collisions may
still be a significant source of high energy emission). If secondary $e^\pm$ were the origin of the radio emission, the
value of $\Gamma$ for protons should be slightly softer than that of secondary $e^\pm$ (Kelner et al. 2006), and the radio
contribution from primary electrons should be minor.  

\subsubsection{The emission in IRAS-N}

Given the values of $n_{\rm c}$ and $v_{\rm bs}$ in IRAS-N, it seems unlikely that the bow shock is accelerating these
electrons. This shock will be strongly radiative, peaking the thermal emission in the optical/UV, with a luminosity of few
times $10^{34}$~erg~s$^{-1}$. On the other hand, the reverse shock is marginally radiative because $\l_{\rm cool}\sim R_{\rm
j}$, peaking at 0.5~keV with a bolometric luminosity $\sim 10^{35}$~erg~s$^{-1}$. As mentioned above, the photoelectric
absorption $\tau_{\rm X}$ is very large, likely $>100$, implying an almost complete suppression of X-rays, which would
explain the non-detection of the source. Concerning the non-thermal radiation, from 
$f_{\rm nt0.1}^{e,p}=1$ one obtains $L_{\rm nt}^{e,p}\sim 5\times 10^{34}$~erg~s$^{-1}$. 

If primary electrons produced the radio emission, the magnetic field should be $B\sim 0.1$~mG for the adopted $f_{\rm nt}$,
and the maximum energy, limited by diffusive escape, $E_{\rm max}\sim 4\times 10^2$~GeV for both electrons and protons. If
secondary $e^\pm$  produced the radio emission, $B> 0.5$~mG, and $E_{\rm max}> 2\times 10^3$~GeV (only protons), limited
again by diffusive escape. 

The density downstream the bow shock may be up to $10^4$ times larger than downstream the reverse shock due to strong
radiative cooling in the former region. Since $t_{\rm RT}$ should be much shorter than $t_{\rm life}$, we assume that the RT
instabilities have time to develop mixing reverse and bow shock downstream material, effectively increasing the density in
the reverse shock. We have adopted a value for $F$ such as that the resulting luminosities due to relativistic Bremsstrahlung
and $pp$ collisions be of the order of $L_{\rm nt}$. In this way, $F$ is used as a free parameter that is optimized to get
high gamma-ray luminosities, not to suppress acceleration, and accounting for the constraints given by the observational data
at lower energies and the $F$-limits provided in Sect.~\ref{posts}. Following such an approach we have adopted $F\sim 25$.
Note that, despite the high density, the Tsytovich-Razin effect may be neglected, since the entrained dense bow shock
material should be cold and hardly fully ionized.

\subsubsection{The emission in HH~80}

In the case of HH~80, both the reverse and the bow shock are adiabatic. We predict thermal X-rays peaking at $\sim 0.5$~keV
with intrinsic luminosities of the order $\sim (R_{\rm j}/l_{\rm cool})\,L_{\rm s}\sim 10^{34}$~erg~s$^{-1}$. This value is
$\sim 100$ times bigger than that given by PTM04, where no intrinsic absorption was assumed. However, if an intrinsic $N_{\rm
H}\sim 5\times 10^{21}$~cm$^{-2}$ in the surroundings of HH~80 were adopted, the factor of 100 could be explained with
photoelectric absorption (see the strong dependence of $l_{\rm cool}$ and $\tau_{\rm X}$ on photon energy and $v_{\rm
r,bs}$). To compute the non-thermal emission, as noted, the two shocks are treated as just one with $f_{\rm nt0.1}^{e,p}=1$,
i.e $L_{\rm nt}^{e,p}\sim 2\times 10^{35}$~erg~s$^{-1}$. 

In the case dominated by primary particles, the magnetic field would be $B\sim 0.005$~mG for the adopted $f_{\rm nt}$, with a
maximum energy limited by diffusive escape $E_{\rm max}\sim 60$~GeV for both electrons and protons. In the case of dominance
by secondaries, $B>0.02$~mG, and $E_{\rm max}>3\times 10^2$~GeV (only protons), limited again by diffusive escape. Since the
medium is quite diluted, relativistic Bremsstrahlung and $pp$ collisions are not as efficient as in IRAS-N, but the fact
that $L_{\rm nt}^{e,p}$ and $t_{\rm esc}$ are both larger renders not so different values for $L_{\rm brems,pp}$.  

\subsection{Computed spectral energy distributions}\label{res}

\begin{figure}
\includegraphics[angle=0, width=0.47\textwidth]{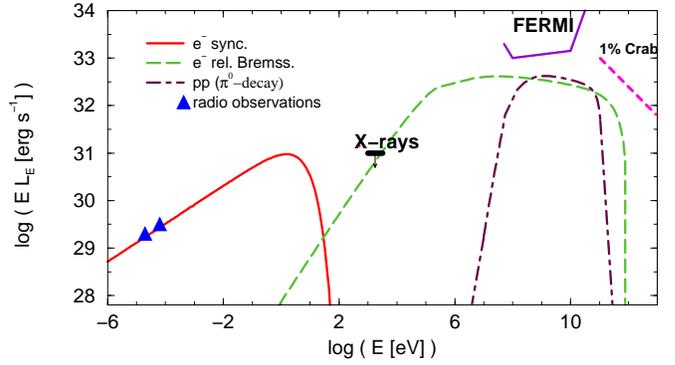}
\caption{Spectral energy distribution of the non-thermal emission for IRAS-N in the primary electron scenario. 
The IC contribution is negligible and not shown here. 
Observational points are from IRAS~16547$-$4247 (radio, Rodr{\'\i}guez et al. 2005;
X-rays, ARA07). The 1~yr/5~$\sigma$ sensitivity of {\it Fermi} in the direction of the galactic 
plane is shown. A curve above 100~GeV showing a luminosity corresponding to 0.01~Crab, 
typical sensitivity of a Cerenkov telescope for exposures of $\sim 50$~hr, is also presented.}\label{SED1}
\end{figure}

\begin{figure}
\includegraphics[angle=0, width=0.47\textwidth]{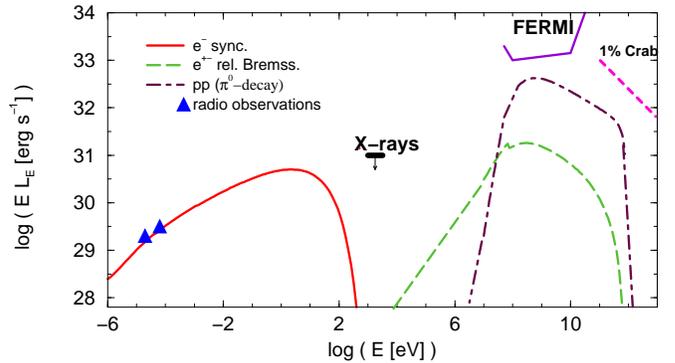}
\caption{The same as in Fig.~\ref{SED1} but for the secondary $e^\pm$ case.}\label{SED2}
\end{figure}

\begin{figure}
\includegraphics[angle=0, width=0.47\textwidth]{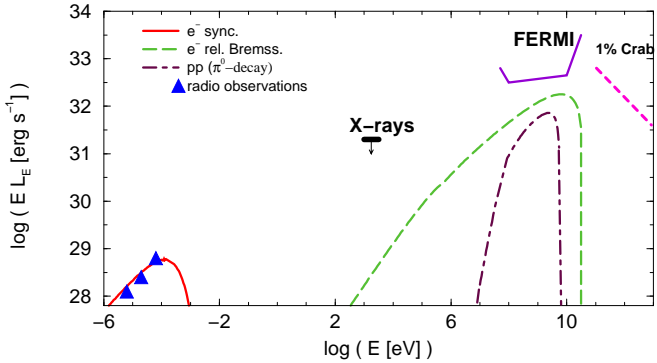}
\caption{The same as in Fig.~\ref{SED1} but for HH~80.
Observational points are from Mart{\'\i} et al. (1993) (radio) and 
PTM04 (X-rays). The X-ray detected point is shown as an upper-limit.}\label{SED3}
\end{figure}

\begin{figure}
\includegraphics[angle=0, width=0.47\textwidth]{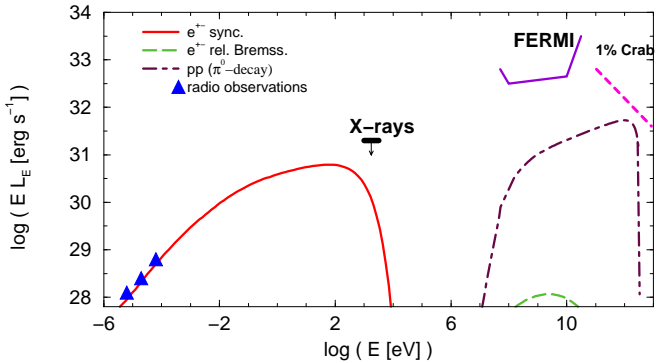}
\caption{The same as in Fig.~\ref{SED2} but for HH~80.}\label{SED4}
\end{figure}

In Figs.~\ref{SED1}, \ref{SED2}, \ref{SED3} and \ref{SED4}, the SEDs computed for IRAS-N and HH~80 are shown. Two scenarios
are adopted for both sources, one in which the radio emission is dominated by primary electrons, and another one in which the
dominant radio emitters are secondary $e^{\pm}$. In the former, $L_{\rm nt}^e$ and $L_{\rm nt}^p$ has been taken equal 
yielding $a\sim 1$; in the latter, we have derived just a lower-limit for $a$ to avoid primary emission to be
significant. In IRAS-N, if secondary $e^\pm$ were the source of radio emission, $a$ should be $>10$. In HH~80, given
the relatively low densities and high magnetic fields of the secondary $e^\pm$ scenario, the proton to electron number
ratio $a$ should be $>1000$. 

It is remarkable that the high-energy components in the SEDs, associated with relativistic Bremsstrahlung and/or $pp$
collisions, have significant luminosities in the high-energy and very high-energy range and fulfill the X-ray constraints.
The synchrotron emission peaks in the optical/UV, and can be the dominant cooling channel of electrons only if the magnetic
field is rather high and densities low. Interestingly, in the primary electron scenario of HH~80, the electron component does
not achieve energies beyond those to emit radio synchrotron emission, but relativistic Bremsstrahlung and $pp$ emission may
still be significant at GeV energies.

The list of the parameter values adopted to calculate the SEDs, together with the radio properties of the sources, is
presented in Table~1. The parameter values have been adjusted numerically and are slightly different from those given in
Sect.~\ref{est}.

 \begin{table*}[]
  \begin{center}
  \caption[]{Radio data and parameters of the non-thermal emitter}
  \label{tab}
  \begin{tabular}{lllll}
  \hline\noalign{\smallskip}
  \hline\noalign{\smallskip}
  & ~~~~~~~~~~~~~~IRAS-N    & &~~~~~~~~~~~~~HH~80    \\
  \hline\noalign{\smallskip}
$\alpha$ & ~~~~~~~~~~~~~~0.5    & &~~~~~~~~~~~~~0.3    \\
$F_{\rm 8~GHz}$ [mJy] & ~~~~~~~~~~~~~~8.7    & &~~~~~~~~~~~~~3    \\
  \hline\noalign{\smallskip}
  & primaries$^{a}$ & secondaries$^{b}$ & primaries & secondaries \\
  \hline\noalign{\smallskip}
$a$ & 1 & $> 10$ & 1 & $> 1000$ \\
$n$ [cm$^{-3}$] & $5\times 10^5$ & $5\times 10^5$ & $1.6\times 10^3$ & $1.6\times 10^3$ \\
$t_{\rm esc}$ [s] & $9\times 10^8$ & $9\times 10^8$ & $4\times 10^9$ & $4\times 10^{11}$ \\
$L_{\rm nt}^{e,p}$ [erg~s$^{-1}$] & $5\times 10^{34}$ & $5\times 10^{34}$ & $2\times 10^{35}$ & $2\times 10^{35}$ \\
$B$ [mG] & 0.25 & 2 & 0.003 & 2.5 \\
$E_{\rm max}$ ($e$) [GeV] & $7\times 10^2$ & $\sim 5\times 10^2$ & 35 & $\sim 10^3$ \\ 
$E_{\rm max}$ ($p$) [GeV] & $7\times 10^2$ & $5\times 10^3$ & 35 & $10^4$ \\ 
$\Gamma$ & 2.2 & 2.4 & 1.6 & 1.8 \\
  \noalign{\smallskip}\hline
  \end{tabular}
  \end{center}
  {
$^{a}$ Primary electron scenario.\\
$^{b}$ Secondary $e^\pm$ scenario.
} 
\end{table*}

\section{Detectability}

The SEDs presented in Figs.~\ref{SED1}-\ref{SED4} show that massive YSOs can produce significant amounts of gamma rays,
although the results are quite sensitive to the available densities downstream. For sources with high densities like IRAS-N,
if the development of RT instabilities does not interfere with particle acceleration, relativistic Bremsstrahlung and $pp$
collisions will be quite efficient. For low-density sources like HH~80, the efficiency of relativistic Bremsstrahlung and
$pp$ collisions is lower, but the lobe sizes are expected to be large, increasing the escape timescales and thereby the
radiation outcome (see Sect.~\ref{lum}). Therefore, for $f_{\rm nt}^{e,p}\ga 0.1$, it can be expected that the termination regions
of massive YSO jets will be eventually detected by Fermi and also by Cherenkov telescopes through long enough exposures. 

In our calculations, the magnetic field strength $B$ and $L_{\rm nt}^{e,p}$ have been adjusted for both to explain the radio
fluxes and to obtain significant gamma-ray fluxes. The magnetic field assumed for high-density, IRAS-N-like sources is in
accordance with estimates derived through Zeeman measurements (e.g. Crutcher 1999). In the case of low-density, HH~80-like,
sources, the situation may be more complicated. 

The primary electron scenario of HH~80 requires a very low magnetic field in the shocked regions. As noted, low densities
imply large lobes and therefore lower magnetic fields in the jet head. However, the value of $B$ in the bow shock can hardly
be smaller than that in the cloud, which is expected to be,  given the cloud densities, several times higher than the one
adopted in our calculations. Therefore, in such a scenario and source type, if detected, gamma rays would likely come from
the reverse shock. Otherwise, in the secondary $e^\pm$ scenario in HH~80 the magnetic field must be quite high, $\sim
2.5$~mG, regardless the shock involved, below but close to the maximum value (see Eq.~\ref{equip}). Furthermore, the value
for $a$ required in this case, $\ga 1000$, may be too large. Also, the hard particle energy distribution required may be
difficult to explain in the context of linear theory of Fermi~I particle acceleration. Such a hard radio spectrum may be
explained by marginal free-free absorption or by an additional thermal component, but then the expected non-thermal fluxes at
higher energies would be smaller due to a softer particle energy distribution. In any case, despite these caveats, one cannot
still rule out HH~80 and similar objects as gamma-ray emitters. 

It is worth noting that the assumptions adopted in this work are quite conservative. The parameter uncertainties are
relatively large, and a more optimistic, but yet consistent with observations, choice of densities, shock velocities and jet
luminosities, could easily move the SED curves up by a factor of several. 

\section{Discussion}

Romero (2008) pointed out that the detection of massive protostars at gamma-ray energies would open a new window to star
formation studies. The detection of the cutoff in the SED would give important insights on the acceleration efficiency in the
terminal shocks of the outflows. The SED can also shed light on the densities, magnetic fields, velocities, and diffusion
coefficients in the shocked regions. Although we do not expect that massive protostars should be among the bright sources
detected by {\it Fermi} (Abdo et al. 2009), our calculations show that they could show up in further analysis of weaker
sources after few years of observation. The emission levels above 100~GeV, close to 0.01~Crab, could be detectable by current
and future Cherenkov telescopes for observation times moderately longer than 50~hours.

However, not only massive protostars, but also the regions in which they form, may be gamma-ray emitters. As mentioned in
Sect.~\ref{mod}, some amount of the highest energy particles may escape to the cloud far upstream the bow shock. It is hard
to estimate the fraction of electrons and protons that would be released in the cloud, which depends strongly on the
diffusion coefficient of the pre-shock cloud medium and the bow shock velocity and size. However, they might carry a
non-negligible fraction of $L_{\rm nt}^{e,p}$ if $\Gamma\sim 2$. In fact, in the case that $L_{\rm nt}^{e,p}\sim 0.01\,L_{\rm
j}$ were in the cloud, massive YSOs may inject an amount of protons well above the average galactic level at several hundreds
of GeV, and the radiation resulting from $pp$ may be detectable (for a general case, see Aharonian \& Atoyan 1996), competing
with that produced in the lobe itself. For leptons, the emission at high energies may be relevant for low magnetic fields,
i.e. when the maximum energy is determined by diffusive escape and dominant relativistic Bremsstrahlung in the cloud. The
spectrum of the gamma rays, generated by $pp$ collisions for protons and relativistic Bremsstrahlung for electrons, should be
very hard since only the highest energy particles escape, peaking at $\epsilon\la E_{\rm max}$. The cloud synchrotron
emission should be quite diluted and dominated by the lobe. 

A clustering of gamma-ray sources should be present in regions with large molecular clouds and star formation, as already
inferred from EGRET data (e.g. Romero et al. 1999). The accumulation of cosmic rays accelerated in the radio lobes into
the molecular cloud can produce extended gamma-ray sources. These radio lobes may be difficult to detect. Neither UV nor hard
X-ray counterparts related to thermal Bremsstrahlung produced in the shock downstream regions are expected to be observed
from these sources because of the large absorption and/or low emission levels. Deep inside the cloud, even radio emission may
be missing due to strong free-free absorption, so the exact number of accelerators could be hard to estimate. Also,
cosmic-ray re-acceleration inside the clouds due to magnetic turbulence (e.g. Dogiel et al. 2004) could result in stronger
sources. Therefore, the combined effect of several protostars deeply embedded in giant clouds might be responsible for
GeV-TeV sources found in star forming regions by EGRET, {\it Fermi}, {\it AGILE} and Cherenkov telescopes. We conclude that
massive clouds with high IR luminosities and maser emission (tracers of massive star formation) deserve detailed study with
{\it Fermi} and ground-based Cherenkov telescopes.    

\begin{acknowledgements}
We thank an anonymous referee for his/her constructive and useful comments that helped to improve the manuscript.
V.B-R., G.E.R., and J.M.P  acknowledge support by the Ministerio de Educaci\'on y Ciencia (Spain) 
under grant AYA 2007-68034-C03-01, FEDER funds. 
V.B-R. wants to thank the Insituto Argentino de Astronom\'ia, and the Facultad de Ciencias Astron\'omicas y 
Geof\'isicas de la Universidad de La Plata, for their kind hospitality.
G.E.R. and A.T.A. are supported by CONICET and the Argentine agency ANPCyT (grant BID 1728/OC-AR PICT 2007-00848).
A.T.A. thanks Max Planck Institut fuer Kernphysik for kind hospitality 
and support.
\end{acknowledgements}

{}
\end{document}